\documentclass[aps,prl,twocolumn]{revtex4}

\usepackage{amsmath}
\usepackage{amssymb,graphicx,array}
\usepackage[hypertex]{hyperref}

\begin{document}
\title{The nuclear symmetry energy and stability of matter in neutron star}
\author{Sebastian Kubis}
\affiliation{Henryk Niewodnicza\'nski Institute of Nuclear Physics,\\
Polish Academy of Sciences,\\
ul.Radzikowskiego 152, 31-342 Krak\'ow, Poland}
\date{\today}
\begin{abstract}
It is shown that behavior of the nuclear symmetry energy is the key 
quantity in the stability consideration in neutron star matter. The symmetry
energy controls the position of crust-core transition and also may lead to
new effects in the inner core of neutron star.
\end{abstract}
\maketitle
\def\ep{\varepsilon}
\def\pa{\partial}
\def\th{\theta}
\def\Kappa{{\cal K}}
\def\ni{\noindent}
\def\Tr{{\rm Tr}}
\def\fm3{\rm ~fm^{-3}}
\def\MeV{\rm ~MeV}
\def\K0{\bar{K}^0}
\def\k0{\bar{k}^0}
\def\ee{{\rm e}}
\def\sp{s^\prime}
\def\anu{\bar{\nu}{}}
\def\dga{\Delta g\,}
\def\coschh{\cos{\chi\over 2}}
\def\gammc#1#2{\bar{#1}\,\Gamma^\mu_{#1#2}\, #2}
\def\gamms#1#2{\bar{#1}\,\tilde{\Gamma}^\mu_{#1#2}\, #2}  
\newcommand{\chih}{\frac{\chi}{2}}
\newcommand{\beq}{\begin{equation}}
\newcommand{\eeq}{\end{equation}}
\newcommand{\beqa}{\begin{eqnarray}}
\newcommand{\eeqa}{\end{eqnarray}}
\newcommand{\arl}{\begin{array}{l}}
\newcommand{\earl}{\end{array}}
\newcommand{\fract}[2]{{\textstyle\frac{#1}{#2}}}
\newcommand{\fracd}[2]{{\displaystyle\frac{#1}{#2}}}
\newcommand{\thh}{\frac{\theta}{2}}
\newcommand{\oht}{\textstyle{\frac{1}{2}}}
\newcommand{\derc}[3]
{\left( \frac{\partial #1}{\partial #2} \right)
 \raisebox{-1em}{\ensuremath{#3}}}
\newcommand{\derct}[3]
{({\partial #1}/{\partial #2})_{#3}}
\newcommand{\npl}{{\it npl }}
\newcommand{\nlep}{n_{lep}}
\newcommand{\enuc}{\ep_{N}}
\newcommand{\elep}{\ep_{L}}

\section{Introduction}
In neutron star, the prevailing part of its interior is fulfilled with  matter
which is in the state called the beta equilibrium \cite{Shapiro:1983du}.
Already few meters
below neutron star surface, at densities $\sim 10^7 g/cm^3$ the degenerated
electrons become relativistic and may easily penetrate the nuclei volume.
The nucleons, although being confined to nucleus, are subject to the beta equilibrium,
e.i. they chemical potentials fulfill the constraint 
$
\mu_n-\mu_p=\mu_e . 
\label{beta}
$
This leads to more and more neutron-rich nuclei and at the density 
$\rho_{drip} \sim 10^{11} g/cm^3$ the {neutron  drip} takes place.
Above the neutron drip density one may
consider the neutron star matter as a two-phase  system in {\em partial} phase
equilibrium. It is, indeed, because not all Gibbs conditions are fulfilled:
$ 
P^d=P^{nuc} \label{gibbs-mech}~,~
\mu_e^d=\mu_e^{nuc} \label{gibbs-e}~,~
\mu_n^{d}=\mu_n^{nuc} \label{gibbs-n}~ 
$
but
$
\mu_p^{d}\neq\mu_p^{nuc}
\label{gibbs-p}~,
$
where superscripts correspond respectively to the dripped and nuclear phase.
At higher density, close to the saturation density the proton drip
occurs as well \cite{pethick}. Above this point $\mu_p^{d}=\mu_p^{nuc}$ and one get a real 
two-phase system with all
Gibbs conditions satisfied. The two phases
have different properties like baryon density $\rho$ and proton fraction $x$. 
Soon after the proton drip the differences in $\rho$ and $x$ vanishes and
the star
matter represents "npl matter" - a homogeneous nucleon-lepton liquid.
(muons start to be produced at slightly higher densities). In the two-phase
system, the nuclei form a lattice leading to a matter with solid state
properties, unlike the homogeneous system being a fluid. In this way  a neutron
star has dense, liquid core covered by solid crust. The inclusion of finite-size
effects, in some models \cite{lorenz-crust}, leads to the presence of "funny
phase" (rods, plates) in  the transition region between
the crust and core. In this short letter we limit ourselves  to the bulk
approximation and shall find the critical density where homogeneous {\it npl}
 matter splits into two phases not going  into details how the phase coexistence
looks like.
In this way we show that the crust-core transition is directly connected  to the
behavior of the nuclear symmetry energy. The same
analysis applied to higher density region reveal the same kind of instability
and suggests repeated solidification in the central part of neutron star core.

\section{Stability conditions}
The beta reactions, taking place in neutron star, conserve charge $Q$ and baryon
number $B$. Such statement means that any thermodynamical state in a volume $V$ 
is uniquely described by setting the temperature $T$ and all conserved quantity,
 here $B$ and $Q$. We neglect the temperature effects which are relevant only
 for a very hot, proto-neutron star. Then the total
energy $U$ becomes a function of volume and conserved numbers $U(V,B,Q)$. In
order to consider stability of single phase one need to introduce intense
(local) quantity $u=U/B$. The energy 
per particle $u$ becomes then a function of other local quantities, taken per
baryon number $v=V/B$ and $q=Q/B$. The first principle of thermodynamics takes
the following  form
\beq
du=-P\;dv-\mu\;dq
\label{first}
\eeq
where $P$ is the pressure and $\mu$ chemical potential of electric charge. From
beta equilibrium one may reads that
\beq
\mu=\mu_e=\mu_\mu=\mu_n - \mu_p
\label{beta-all}
\eeq
The minus sign before $\mu$ in (\ref{first}) comes from the definition of
$Q=N_p-N_e-N_\mu$ which is negative for leptons \footnote{this convention of
charge sign is opposite to that used in \cite{Kubis:2004xh} and of course more 
natural}. The stability of any single phase, also called the {\em intrinsic}
stability, is ensured by the convexity of $u(v,q)$ \cite{callen}.
Thermodynamical identities allows to express this requirement in terms
of following inequalities \cite{Kubis:2004xh}

\beq
-\derc{P}{v}{q} > 0~~~~~~ -\derc{\mu}{q}{P} > 0
\label{posit1} 
\eeq
Usually, only the positive compressibility is examined, in particular, it is 
required for locally neutral matter that
\beq -\derc{P}{v}{q=0} > 0. \eeq
However the second inequality in (\ref{posit1})  is of the
same  importance. It concerns the stability of charge fluctuations and as it
will be shown later  it is connected  to the positive value of the screening
length in matter.
Not all nuclear models ensure the charge fluctuations to be stable. As was shown
in the case of kaon condensation for wide class of models  the system is  not
stable at any density \cite{Kubis:2004xh}. In this work we would like to show
that even such simple system like $npl$ mater, without any exotic components,
may represent a region of density where instability occurs.

One may find another pair of inequalities which are equivalent to those in 
(\ref{posit1}) and, as it will appear later,
are more convenient in further calculations:
\beq
-\derc{P}{v}{\mu} >0 ~~~~~~ -\derc{\mu}{q}{v} > 0 .
\label{posit2}
\eeq
The intrinsic stability is determined by the details of nucleon-nucleon
interactions. 
In order to show that, let's split the total
energy per baryon into the nucleonic and leptonic part
\beq
u = u^N + u^L~.
\eeq
The nucleonic contribution may be always expressed  as a function of baryon number
density $n=B/V$ and the proton fraction $x=N_p/B$.
For leptons $\ep^L$, the 
energy per volume, is completely determined by their chemical potential
$\mu$. Such decomposition is also true for the
total pressure, so one may write
\beqa
u &=& u^N(n,x) + \ep^L(\mu)/n \\
P &=& P^N(n,x) + P^L(\mu)
\eeqa
where $P^N = n^2 u^N_n$, henceforth, the partial derivatives of $u^N$ or $P^N$
we indicate by subscripts $n$ or $x$.  
One may show that $\mu_n-\mu_p = \derct{u^N}{x}{n}$
and then the beta equilibrium means that 
\beq
\mu = - u^N_x.
\label{beta-short}
\eeq
Below we show that the equation (\ref{beta-short})
allows to express the stability conditions  (\ref{posit1}) or (\ref{posit2}) in
terms only of the nucleonic contribution  $u^N$ to the total energy $u$.
First, we take the compressibility from the second pair of inequalities 
(\ref{posit2})
\beqa
-\derc{P}{v}{\mu}&=&n^2\derc{P^N}{n}{\mu} = \\
&=&n^2\left(P^N_{n}+P^N_{x}\derc{x}{n}{\mu}\right)
\label{compr-mu}
\eeqa
the leptons does not contribute when the derivative is taken under $\mu$ fixed.
From (\ref{beta-short}) one get $\derct{x}{n}{\mu} = -{u^N_{nx}}/{u^N_{xx}}$
and joining with  (\ref{compr-mu}) we obtain
\beq
- \derc{P}{v}{\mu} = n^4\left(2\;\frac{u^N_n}{n} + u^N_{nn} -
\frac{(u^N_{nx})^2}{u^N_{xx}}\right) 
\label{compr-unx}
\eeq
where for the right-hand side we applied the expression for the nucleonic
pressure $P^N = n^2 u^N_n$. The first two terms in the (\ref{compr-unx}) refers
to the nucleonic pressure and compressibility and they are positive from very
fundamental reasons. As we show later, the third term may lead to negative
contribution and  it comes from the leptons presence, it is the leptons that
make the matter unstable.

In order to find similar expression for the second inequality in (\ref{posit2})
we make use of the expression for the total charge per baryon
$
q= x - n_L/n
$~,
where the lepton number density $n_L$ depends on $\mu$ only (see Appendix).
Differentiation of the equation by $\mu$ under constant $n$ we obtain
\beq
\derc{q}{\mu}{n} = \derc{x}{\mu}{n} - \frac{n_L'(\mu)}{n}
\label{qmun}
\eeq
Here one may note that the second stability condition in (\ref{posit2}) is 
directly connected to the stable screening of charged particles. The derivative
$n_L'$ appearing in (\ref{qmun}), is proportional to the sum of squared
inverse  of screening lengths for leptons \cite{Heiselberg:1992dx}.  The first
term in (\ref{qmun}), however, lacks this simple interpretation, it refers to
hadronic interactions concealed by the beta equilibrium constraint.  It is
useful to call quantities like $\pa q/\pa \mu$, as the
"electrical capacitance" of matter. It measures the energetic cost of change
in electric charge held in matter. For further discussion we introduce the 
compressibility and electric capacitance as
\beqa
K_i &= &-v^2\derc{P}{v}{i}=\derc{P}{n}{i}~,~~~i=q,\mu\\
\chi_j& =& -\derc{q}{\mu}{j}~,~~~j=P,v
\eeqa
By use of the beta equilibrium relation (\ref{beta-short})  one may find that
$(\pa x/\pa {\mu})_{n} = -1/u^N_{xx}$ what allows to write the conditions
(\ref{posit2}) only  in terms of nucleonic energy  contribution  $u^N$
\beqa
K_\mu& =& 2\, n\, {u^N_n}  + n^2 u^N_{nn} -  \frac{(u^N_{nx}n)^2}{u^N_{xx}} >0 
\label{Kmu+} \\
\chi_v& =& \frac{1}{u^N_{xx}} + \frac{\mu(k_e+k_\mu)}{n\pi^2} > 0 .
\label{chiv+}
\eeqa
because also the second term in (\ref{chiv+}), coming from leptons, is
determined by the relation $\mu=-u^N_x$ and nucleonic pressure is $P^N=n^2u^N_n$.
In the same way we may express the first
pair of conditions (\ref{posit1}). In order to derive them it is useful to
find identities between  compressibilities and capacities. Then using standard
theorem for implicit functions one may get the following relations
\beqa 
\chi_P-\chi_v&=& K_\mu\alpha_P^{-2}n^{-2}  \label{chiP-v}\\
K_q-K_\mu&=&{\chi_v \alpha_q^2}n^2 \label{Kq-mu}\\ 
\frac{K_q}{K_\mu}&=&\frac{\chi_P}{\chi_v} 
\eeqa
with 
\beq
\alpha_i=\derc{\mu}{n}{i}~~~~i=q,P 
\label{alphas}
\eeq
Above relations perform  analogy between the  beta stable matter and the
one-component system at non-zero temperature. For instance, if one replace  the
chemical potential by the temperature, $\mu\rightarrow T$, and the charge by
the entropy $q\rightarrow s$ in the Eq.(\ref{chiP-v}), then recover the well
known relations between specific heats: $C_P-C_V=T(\pa P/\pa V)_T(\pa V/\pa
T)_P^2$. Derivatives defined in (\ref{alphas}) also may be
expressed by means of $u^N$ (see Appendix) so the conditions (\ref{posit1}) 
finally  takes the form
\beqa
K_q& =&n^2 u^N_{nn} + 2nu^N_{n} + \nonumber\\
&&\frac{n_L^2 v^2 u^N_{xx} -
u^N_{nx}(2n_L+n_L'n\;u^N_{nx})}
{1+u^N_{xx}n_L'v} >0 
\label{Kq+} \\
\chi_P&=& \frac{P^N_n + u^N_{xx}n_L^2v^2 - 2 u_{nx}n_L}
{P^N_n u^N_{xx} - (u^N_{nx}n)^2} +\frac{\mu(k_e+k_\mu)}{n\pi^2}  > 0 \nonumber\\
&& \label{chiP+}
\eeqa
For further analysis we choose the stability conditions expressed by the pair
of equations (\ref{Kmu+},\ref{chiv+}). They are not only more readable 
than (\ref{Kq+},\ref{chiP+}) but also more sensitive to the onset of
instability. One may deduce it from
Eq.(\ref{chiP-v},\ref{Kq-mu}). Any stable system always keep
\beq
K_\mu < K_q~~ {\rm and }~~ \chi_v < \chi_P~.
\eeq
and thus the quantities $K_\mu$ or $\chi_v$ vanish first.

As was mentioned above the nucleonic contribution $u^N$ to the total energy, is
a function of $n$ and $x$. Isospin symmetry allows for the expansion in even
powers of $(1-2x)$ but  terms higher than quadratic are negligible, then the 
energy per baryon takes the form
\beq
u^N(n,x) = V(n) + E_s(n)(1-2x)^2~,
\label{uN}
\eeq
where $V(n)$ is the isoscalar potential and $E_s(n)$ - the symmetry energy,
corresponding to the interactions isovector channel. 
Applying (\ref{uN}) to (\ref{Kmu+},\ref{chiv+}) we obtain
\beqa
K_\mu&=&
n^2\left(E_s'' (1-2 x)^2+V''\right) + 
2\; n\!\left(E_s' (1-2 x)^2+V'\right) \nonumber\\
&&-\frac{2 (1-2 x)^2 E_s'^2 n^2}{E_s} > 0
\label{Kmu+Es}\\
\chi_v&=&\frac{1}{8 E_s(n)} + \frac{\mu(k_e+k_\mu)}{n\pi^2}  >0
\label{chiv+Es}
\eeqa
Above equations shows explicitly
 the importance of symmetry energy $E_s$ in the
stability considerations.  The concrete shape of the function 
$E_s(n)$ is not well
known. We know only its value at saturation point, $n_0=0.16~\rm
fm^{-3}$, where it takes about $30$~MeV. Recent analysis on the isospin
diffusion in heavy-ion collisions  constrained significantly the slope of
$E_s'(n_0)$ and the stiffness $E_s''(n_0)$ at saturation point
\cite{Chen:2005ti,Chen:2004si}, however these results does not determine the
high  density behavior definitely. There are 
no experimental evidence about values of $E_s$ at very high density
which is available in the central parts of a neutron star.
In such extrapolations we must rely on the model calculations. For all of them
the symmetry energy at saturation point have positive slope  but at higher
densities, they lead to different conclusions. For most the $E_s$ is 
monotonically  increasing function of $n$ but some models lead to  the $E_s$
which saturates at higher densities or even bends down at some point and goes
to zero \cite{WFF,APR,Kaiser:2001jx}. This kind of behavior is
especially interesting as the last term in (\ref{Kmu+Es}) may then take
arbitrary large negative values and lead to instability. From the other side,
going to very low density, we encounter uncertainties as well.  
All models predict $E_s$ decreasing to 0. 
However recent experiments \cite{Kowalski:2006ju} show that symmetry energy
saturates at the level about 10 MeV for very low densities.
 
\section{Nuclear models}
\begin{figure}[h]
\includegraphics[width=.4\textwidth,clip]{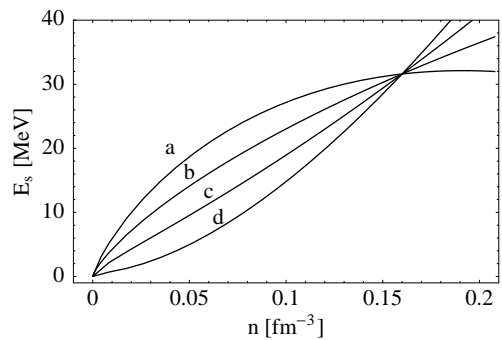}
\caption{The different shapes of the symmetry energy at 
densities below saturation point.}
\label{Ess-fig-low}
\end{figure}

\begin{figure}[h]
\includegraphics[width=.4\textwidth,clip]{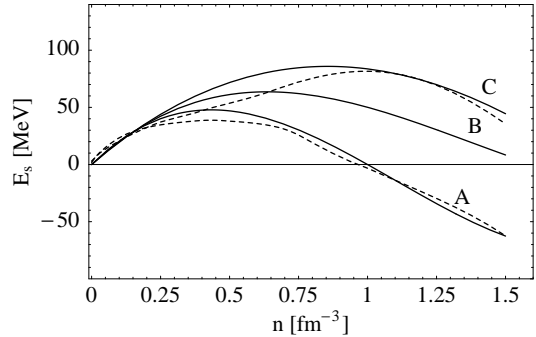}
\caption{Three different shapes of the symmetry energy at 
densities above saturation point (solid lines). For comparison 
the results of realistic potentials (dotted lines), the higher: UV14UVII, 
and the lower: UV14TNI.}
\label{Ess-fig-high}
\end{figure}

In order to present the role played by the symmetry energy we apply a set 
nuclear models.
To emphasize the $E_s$ effects, the isoscalar part $V(n)$ is kept the same,
whereas the symmetry energy takes different forms. 
The isoscalar potential $V(n)$, is that one taken from \cite{Prakash:1988md}
which  leads to  the compressibility of nuclear matter equal to 240 MeV at
saturation point.
For $E_s$ we used two family of functions: four at low density and three 
for high density regime.
At lower densities, we use shapes applied by  Chen et.al. in 
\cite{Chen:2004si}, there were numbered by a parameter $x=1,0,-1,-2$.
Here we named them by a,b,c,d  to avoid confusion with proton
fraction $x$. The shapes of symmetry energy dependence at lower densities
are presented in Fig.\ref{Ess-fig-low}. 
At higher densities, much above $n_0$, we applied a "bent down" function.
This type of symmetry energy  with low values of $E_s$ at high density is
typical for realistic potentials
\cite{WFF,APR}. Also in modern approaches like chiral dynamics
\cite{Kaiser:2001jx} and Skyrme effective forces \cite{Agrawal:2006ea}
small values of $E_s$ were obtained.
Here, for numerical simplicity, we introduced the 
simple polynomial function 
which imitate results of works mentioned above
\beq
E_s(n) = E_s^{(0)}\frac{n(n-n_1)(n-n_2)}{n_0(n_0-n_1)(n_0-n_2)} ,
\eeq
where $E_s^{(0)}\!=\!30$~MeV and $n_i$ are the zeros of
 $E_s$ (see Appendix). 
The shapes of these functions, named A, B, C
are shown in the Fig.\ref{Ess-fig-high}.

\section{Low densities region}
\begin{figure}
\includegraphics[width=.4\textwidth]{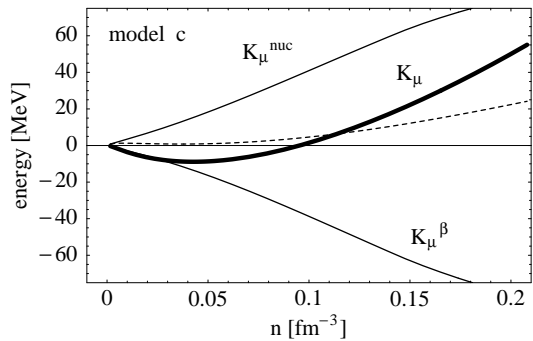}
\caption{The compressibility $K_\mu$ (thick) and its  contributions (thin
lines). The dotted line corresponds to the energy per baryon for neutral matter
 $u(n,0).$}
\label{Kcontrib-fig}
\end{figure}

The transition between liquid core and solid crust of the star is strictly 
connected to the breaking of the conditions (\ref{Kmu+Es}, \ref{chiv+Es}). When
at least one of them is broken the matter cannot be homogeneous anymore, it
splits into two phases. The Fig.\ref{Kcontrib-fig} shows the
compressibility under constant $\mu$  and its two contribution: "nuclear"
$K_\mu^{nuc}$ - the two first terms in  (\ref{Kmu+Es}) - and "beta"
$K_\mu^\beta$ - the last term in (\ref{Kmu+Es}) which comes from the leptons
presence. The "beta" contribution is always
negative, hence there is always a competition between the positive "nuclear"
compressibility and the beta 
reactions which tends to destabilize the  matter. At some critical point, $n_c$,
the compressibility changes its sign and below $n_c$ the matter cannot exist as
a single, neutral phase.
The actual splitting into two phases does not occur exactly at $n_c$, but
slightly above $n_c$,
because the system must find a state where the two charged phases may coexist.
The correction is expected to be tiny, so the critical point for the vanishing of
$K_\mu$ may be treated as a good estimation for the boundary of the liquid core
in neutron star.
The Table \ref{tab1} shows the critical density at which $K_\mu$ vanishes.
It depends strongly on $E_s$ but does not behaves monotonically with the 
$E_s$. For symmetry energy taking both high and low values (a, d case) 
the $n_c$ moves to higher density close to saturation point. The lowest 
values of $n_c$ are achieved with intermediate  $E_s$ (models b and c).
There is no simple relation between values of $E_s$ and $n_c$ because
the first and the second derivatives of $E_s$ are essential as well.

\begin{table}[h]
\caption{The critical density for different models.}
\begin{tabular}{c|cccc}
model & a&b&c&d\\
\hline
$n_c ,\fm3$&~ 0.119~ &	~0.092 ~&~	 0.095~	&~ 0.160 ~\\
\end{tabular}   
\label{tab1}
\end{table} 

\section{High densities region}
\begin{figure}[h]
\includegraphics[width=.4\textwidth]{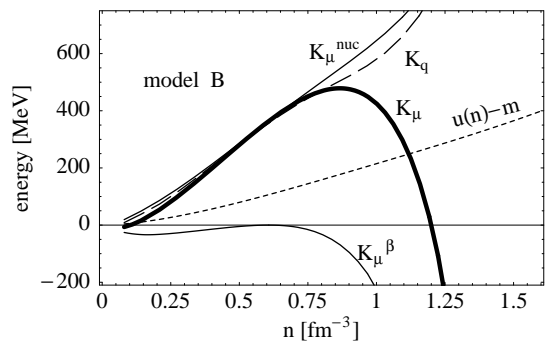}
\caption{The compressibility $K_\mu$ (thick) and its  contributions (thin
lines). The dashed line corresponds to $K_q$ and dotted line
energy per baryon for neutral matter, $u(n)\equiv u(n,0)$}
\label{Kcontrib-high-fig}
\end{figure}

The stability of matter at densities much greater than $n_0$
do not need to be obvious. For the symmetry energy increasing in the whole range
of density the matter is stable indeed. However, 
the chosen nuclear models with 
very low values of $E_s$ lead again to the same kind of instability as occurs
in the crust-core transition region. For all presented models A,B,C
there is a critical density $n_c$ where $K_\mu$ vanishes. The behavior of 
the compressibility $K_\mu$ for the model B is shown on the
Fig.\ref{Kcontrib-high-fig}.  It is worth to notice that energy per baryon
for neutral matter $u(n)\equiv u(n,0)$
 does not reveal any pathology, it is monotonically increasing and 
its convexity, $u_{nn}(n) = K_q/n^2$, remains positive. If one look
only on the energy per baryon behavior, one may overlook that the matter
becomes unstable at some point. 

Of course, the instability point has physical meaning only if it is attainable 
in a neutron star. The Tab.\ref{tab2} shows the neutron star
properties (the central density $n_{cen}$ of a star with maximal mass $M_{max}$
found by solving of TOV equation 
\footnote{The results of TOV equations are presented under assumption
that single phase EOS is stable for all densities up to the max density,
also above $n_c$. It only shows that $n_c$ is attainable
for a given EOS. Above $n_c$ the EOS should be corrected 
and $n_{cen}$ may change but not much.}.
\begin{table}[h]
\caption{The critical density and neutron star properties for different models.} 
\label{tab2}
\begin{tabular}{l|ccc}
model & A&B&C\\
\hline
$n_{cen},\fm3$& 1.92& 1.32&1.21 \\
$M_{max}/M_\odot$& 1.64& 1.73& 1.84\\
$n_c ,\fm3$&~ 0.74~ &~	1.20~ &~ 1.43~ \\
\end{tabular}   
\end{table} 
Actually, in the case A and B the $n_c$ is smaller than the central density
of maximal star.
It means that for sufficiently massive stars 
their structure changes essentially in the central part of the star.
The homogeneous phase again splits into two-phase system. One may suspect  
formation of funny phases or solidification of the central part of stellar core.

\section{Summary}
In this letter we would like to notice the role played by the nuclear energy
symmetry $E_s$ in the stability of dense matter under beta equilibrium.
It was shown that relevant quantity in such considerations is $K_\mu$ - the
compressibility under constant chemical potential, rather than $K_q$ - the
compressibility under constant charge.
The instability of matter leads to phase separations and corresponds to the
transition from the liquid core to the solid crust.
In this way one may get simple 
connection between $E_s$ behavior at low densities and the size of the star
crust, which may be estimated from pulsar glitching \cite{alpar}. 

The stability considerations were also carried out at very high density. 
It was shown that for nuclear models with small values of $E_s$ it  is possible
that instability occurs. This may lead to repeated solidification the
internal parts of a core. 
Such solid inner core would have important consequences on the rotation of 
a star and its magnetic properties.
\appendix
\section{APPENDIX}
Here we present formulas mentioned in the regular text,
the lepton number density
\beq
n_L=\frac{k_e^3}{3\pi^2}+\frac{k_\mu^3}{3\pi^2} = 
\frac{\mu^3}{3\pi^2}+\th(\mu-m_\mu)\frac{(m_\mu^2 -
\mu^2)^{3/2}}{3\pi^2}
\nonumber
\eeq
and mixed derivatives appearing in (\ref{chiP-v}) and (\ref{Kq-mu})
\beqa
\alpha_q&=&\derc{\mu}{n}{q}=\frac{u^N_{xx}n_Lv^2-u^N_{nx}}{1+u^N_{xx}n_L'v}
\nonumber \\
\alpha_P&=&\derc{\mu}{n}{P}=
\frac{u^N_{xx}(n^2u^N_{nn}+2n u^N_n) - (u_{nx}n)^2}
{n^2 u^N_{nx} - n_L u^N_{xx}}
\nonumber
\eeqa
The parameterization of $E_s$ at high density  regime:
\begin{table}[h]
\begin{tabular}{l|ccc}
model & A&B&C\\
\hline
$n_1,\fm3$&~1.0~&~1.5~&~1.8~ \\
$n_2,\fm3$& 2.3& 2.5& 10.0\\
\end{tabular}   
\end{table} 
\bibliography{enstab}
\end{document}